\documentstyle[aps,prl,twocolumn,floats,epsfig]{revtex}

\newcommand{\ket}[1]{\left\vert#1\right\rangle}

\newcommand{\one}{\mbox{$1 \hspace{-1.0mm}  {\bf l}$}}

\begin{document}

\title{Entanglement enhanced information transmission  over
a quantum channel with correlated noise}

\author{Chiara Macchiavello}


\address{Dipartimento di Fisica ``A.Volta", Via Bassi 6,
I-27100 Pavia, Italy \\ and Istituto Nazionale per la Fisica della Materia (INFM)}
\author{G.Massimo Palma}


\address{Dipartimento di Scienze Fisiche ed Astronomiche,  Via
Archirafi 36, I-90123 Palermo, Italy \\
and NEST - Istituto Nazionale per la Fisica della Materia (INFM)}

\date{\today}

\maketitle

\begin{abstract}
{\bf We show that entanglement is a useful resource to enhance the mutual information of the depolarizing channel
when the noise on consecutive uses of the channel has some partial correlations. We obtain a threshold in the
degree of memory, depending on the shrinking factor of the channel, above which a higher amount of classical
information is transmitted  with entangled signals.}
\end{abstract}


The classical capacity of quantum channels, i.e. the amount of classical information which can be reliably
transmitted by quantum states in the presence of a noisy environment has received renewed interest in recent years
\cite{Holevo}. One of the main focuses of such interest is the study of entanglement as a useful resource to
enhance the classical channel capacity. Although the theory does not rule this possibility out, the search for
superaddivity of quantum channels has led sofar to the evidence that no such property is present in memoryless
channels. This has been first proved analytically for the case of two entangled uses of the depolarizing channel
\cite{Bruss} and then extended to a broader class of memoryless channels  \cite{King}. In this paper we will turn
our attention to a different class of channels, namely to channels with partial memory. For such channels our
results show that a higher mutual information can indeed be achieved above a certain memory threshold by entangling
two consecutive uses of the channel. In the following  each use of the channel will be a qubit, i.e will be a
quantum state belonging to a two-dimensional Hilbert space. The action of transmission channels is described by
Kraus operators

\cite{Kraus} $A_i$, satisfying $\sum_i A^{\dagger}_iA_i = \one$, such that if we send through the channel a qubit
in a state described by the density operator $\pi$ the corresponding output state is given by the map

\begin{equation}
\pi \longrightarrow \Phi (\pi ) =\sum_i A_i \pi A_i^{\dagger}
\end{equation}

An interesting class of Kraus operators acting on individual qubits  can be expressed in terms of the Pauli
operators $\sigma_{x,y,z}$

\begin{equation}
A_{i} = \sqrt{p_{i}} \sigma_{i}\ ,
\end{equation}

with $\sum_i p_i = 1$ , $i=0,x,y,z$ and $ \sigma_{0}= \one$. A noise model for these actions is for instance the
application of a random rotation of the angle $\pi$ around axis $ \hat{\bf{x}},\hat{\bf{y}},\hat{\bf{z}}$ with
probability $p_x , p_y , p_z$ and the identity with probability $p_0$.

In the simplest scenario the transmitter can send one qubit at a time along the channel. In  this case the
codewords will be restricted to be the tensor products of the states of the individual qubits. Quantum mechanics
however allows also the possibility to entangle multiple uses of the channel. For this more general strategy it has
been shown that the amount of reliable information which can be transmitted per use of the channel is given by
\cite{Holevo}

\begin{equation}
C_n =\frac{1}{n}{\mbox{sup}}_{\cal E} I_n(\cal E)\ \ ,
\end{equation}

where ${\cal E}=\{P_i,\pi_i\}$  with $P_i\geq 0, \sum P_i=1$ is the input ensemble of states $\pi_i$, transmitted
with {\it a priori} probabilities $P_i$, of  $n$ -- generally entangled -- qubits and $I_n({\cal E})$ is the mutual
information

\begin{equation}
I_n({\cal E}) = S(\rho ) - \sum_i P_i S(\rho_i)\ \ , \label{def-I}
\end{equation}

where the index $n$ stands for the number of uses of the channel. Here

\begin{equation}
S(\chi ) = - {\mbox{tr}}(\chi \log \chi )
\end{equation}

is the von Neumann entropy,  $\rho_i = \Phi (\pi_i)$ are the density matrices describing the output states and
$\rho = \sum_i P_i \rho_i$. Logarithms are taken to base 2. The advantage of the expression \ref{def-I}) is that it
includes an optimization over all possible POVMs at the output, including collective ones. Therefore no explicit
maximization procedure for the decoding at the output of the channel is needed.

The interest for the possibility of using entangled states as channel inputs is motivated by the fact that it
cannot generally be excluded that $I_n({\cal E})$ is  superadditive in the presence of entanglement, i.e. we might
have $I_{n+m}> I_n + I_m$ and therefore $C_n > C_1$.

In this scenario the classical capacity $C$ of the channel is defined as

\begin{equation}
C=\lim_{n\to \infty}  C_n\;. \label{capac}
\end{equation}

Sofar the main objects of investigation have been memoryless channels. By definition a channel is memoryless when
its action on arbitrary signals $\pi_s$, consisting of $n$ qubits (including entangled ones), is given by

\begin{equation}
\Phi (\pi_s ) = \sum_{i_1\cdots i_n} (A_{i_n}\otimes \cdots \otimes A_{i_1}) \pi_s (A^{\dagger}_{i_1}\otimes \cdots
\otimes A^{\dagger}_{i_n})
\end{equation}

In the case of Pauli channels a more general situation is described by action operators of the following form

\begin{equation}
A_{k_1\dots k_n} = \sqrt{p_{k_1\dots k_n}} \sigma_{k_1} \dots \sigma_{k_n}\;,
\end{equation}

with $\sum_{k_1\dots k_n} p_{k_1\dots k_n} =1$. The quantity $ p_{k_1\dots k_n}$ can be interpreted as the
probability that a given random sequence of rotations of an angle $\pi$ along axis $k_1\dots k_n$ is applied to the
sequence of $n$ qubits sent through the channel. For a memoryless channel  $p_{k_1\dots k_n}=  p_{k_1}p_{k_2}\dots
p_{k_n}$. An interesting generalization is described by a Markov chain defined as

\begin{equation}
 p_{k_1\dots k_n} =  p_{k_1} p_{k_2|k_1}\dots  p_{k_n|k_{n-1}}
\end{equation}

where $p_{k_n|k_{n-1}}$ can be interpreted as the conditional probability that a $\pi$ rotation around axis $k_n$
is applied to the $n$-th qubit given that a $\pi$ rotation around axis $k_{n-1}$  was applied on the  $n-1$-th
qubit. Here we will consider the case of two consecutive uses of a channel with partial memory, i.e. we will assume
$ p_{k_n|k_{n-1}} =   (1-\mu ) p_{k_n} + \mu \delta_{k_n|k_{n-1}}$. This means that with probability $ \mu $ the
same rotation is applied to both qubits while with probability $1 - \mu $ the two rotations are uncorrelated.

In our noise model the degree of memory $\mu$ could depend on the time lap between the two channel uses. If the two
qubits are sent at a very short time interval the properties of the channel, which determine the direction of the
random rotations, will be unchanged, and it is therefore reasonable to assume that the action on both qubits will
take the form

\begin{equation}
A^{c}_{k} = \sqrt{p_{k}} \sigma_{k} \sigma_{k}\;.
\end{equation}

If on the other hand, the time interval between the channel uses is such that  the channel properties have changed
then the actions will be

\begin{equation}
A^{u}_{k_1,k_2} = \sqrt{p_{k_1}}\sqrt{p_{k_2}} \sigma_{k_1}\sigma_{k_2}\;.
\end{equation}

An intermediate case, as mentioned above, is described by actions of the form

\begin{equation}
A^{i}_{k_1,k_2} = \sqrt{ (1-\mu ) p_{k_n} + \mu \delta_{k_n|k_{n-1}}  } \sigma_{k_2}\sigma_{k_2}\;. \label{part}
\end{equation}

It is straightforward to verify that the Bell states, defined in the basis $\ket{0},\ket{1}$ of the eigenstates of
the $\sigma_z$ operators as

\begin{eqnarray}
\ket{\Phi_{\pm}} & = &\frac{1}{\sqrt 2} \{ \ket{00} \pm \ket{11}\}
\nonumber\\
\ket{\Psi_{\pm}} & = &\frac{1}{\sqrt 2} \{ \ket{01} \pm \ket{10}\}
\end{eqnarray}

are eigenstates of the operators $A^{c}_{k}$ and therefore will pass undisturbed through the channel. If used as
equiprobable signal states they maximise $I_2$, as we will have $I_2 = 2$. Furthermore it is immediate to verify
that the value $I_2 = 2$ cannot be achieved by any ensemble of tensor product input states. This situation is
reminiscent of the so called noiseless codes, where collective states are used to encode and protect quantum
information against collective noise \cite{Palma}.

In the following we will concentrate our attention to the depolarizing channel, for which $p_0=1-p$ and $p_i=p/3,
i=x,y,z$. We will consider an ensemble of orthogonal input states parametrised as follows

\begin{eqnarray}
|\pi_{1}\rangle & = & \cos\vartheta |00\rangle + \sin\vartheta
|11\rangle\nonumber\\
|\pi_{2}\rangle & = & \sin\vartheta |00\rangle -  \cos\vartheta
|11\rangle\nonumber\\
|\pi_{3}\rangle & = & \cos\vartheta |01\rangle + \sin\vartheta |10\rangle\nonumber\\
|\pi_{4}\rangle & = & \sin\vartheta |01\rangle - \cos\vartheta |10\rangle \;. \label{signals}
\end{eqnarray}

Although it is not a priori certain that this is the optimal choice for all values of $\mu$  we know that it
maximizes $C_2$  with $\vartheta = 0 $ for $\mu = 0$ (uncorrelated noise), and with $\vartheta = \frac{\pi}{4} $
for $\mu = 1$ (fully correlated noise). We will therefore optimize the ansatz (\ref{signals}) by looking for the
value $\vartheta (\mu )$ which maximizes $I_2$ as a function of $\mu$.

We will now show that there is a  threshold value  $\mu_t $ for which $I_2(\vartheta = \frac{\pi}{4},\mu_t ) = I_2(
\vartheta = 0 ,\mu_t )$. Below the threshold value $I_2(\vartheta=0,\mu <\mu_t)
>I_2(\vartheta=\frac{\pi}{4},\mu <\mu_t)$ while above
$I_2(\vartheta=\frac{\pi}{4},\mu >\mu_t)
>I_2(\vartheta=0,\mu >\mu_t)$.
To this goal it is useful to use the Bloch representation \cite{Mahler} for the states

\begin{eqnarray}
\pi = \frac{1}{4} \left\{ \one\otimes\one + \one\otimes \sum_k\beta_{k}^{(2)} \sigma_{k} + \sum_k\beta_{k}^{(1)}
\sigma_{k} \otimes\one\right.\nonumber\\
\left. +\sum_{kl} \chi_{kl}\sigma_{k}\otimes\sigma_{l} \right\}
\end{eqnarray}

where the Bloch vectors and tensor are defined respectively as $\beta_i = {\mbox{tr}}(\pi \sigma_i ) , \chi_{ij} =
{\mbox{tr}}(\pi \sigma_i\sigma_j )$. We will express the action of the channel in terms of the so called shrinking
factor $\eta=1-4p/3$.

It is straightforward to verify that for $\mu=0$

\begin{eqnarray}
&&\sum_{k_1,k_2}A_{k_1,k_2}\one\otimes\sigma_j A_{k_1,k_2}^\dagger=
\eta\one\otimes\sigma_j\nonumber\\
&&\sum_{k_1,k_2}A_{k_1,k_2}\sigma_j\otimes\one A_{k_1,k_2}^\dagger=
\eta\sigma_j\otimes\one\nonumber\\
&&\sum_{k_1,k_2}A_{k_1,k_2}\sigma_k\otimes\sigma_j A_{k_1,k_2}^\dagger= \eta^2\sigma_k\otimes\sigma_j \label{sm}
\end{eqnarray}

while for $\mu=1$

\begin{eqnarray}
\sum_{k_1,k_2}A_{k_1,k_2}\one\otimes\sigma_j A_{k_1,k_2}^\dagger
&=&\eta\one\otimes\sigma_j\nonumber\\
\sum_{k_1,k_2}A_{k_1,k_2}\sigma_j\otimes\one A_{k_1,k_2}^\dagger
&=&\eta\sigma_j\otimes\one\nonumber\\
\sum_{k_1,k_2}A_{k_1,k_2}\sigma_k\otimes\sigma_j A_{k_1,k_2}^\dagger&=&\nonumber\\
=\delta_{kj}\sigma_k\otimes\sigma_j&+&(1-\delta_{kj})\eta\sigma_k\otimes\sigma_j\;. \label{cm}
\end{eqnarray}

It is interesting to note that both for $\mu=0$ and for $\mu=1$ the components of the Bloch vectors $\beta_k^{(i)}$
of the input states are shrunk isotropically by the shrinking factor $\eta$. The difference between the two cases
is the action on the Bloch tensor $\chi$. The input state $\ket{\pi_1}$ is transformed by the action of the
depolarizing cannel with partial memory defined in equation (\ref{part}) into the output state $\rho_1$

\begin{eqnarray}
\rho_1 &=& \frac{1}{4}\{ \one\otimes\one + \eta\cos 2\vartheta (\one\otimes\sigma_z + \sigma_z\otimes\one ) + \\
&&[\mu + (1- \mu)\eta^2][\sigma_z\otimes \sigma_z + \sin 2\vartheta ( \sigma_x \otimes\sigma_x - \sigma_y\otimes
\sigma_y )] \}\nonumber
\end{eqnarray}

The corresponding eigenvalues are:

\begin{eqnarray}
\lambda_{1,2} &=&\frac{1}{4}(1-\mu)(1-\eta^2)\\
\lambda_{3,4}&=&\left.\frac{1}{4} \right\{ 1+\mu+\eta^2(1-\mu)\nonumber\\
&&\left.\pm 2\sqrt{\eta^2\cos^22\vartheta +[\eta^2(1-\mu)+\mu]^2\sin^22\vartheta}\right\}\ \label{eigenvalues}
\end{eqnarray}

Notice that the first two eigenvalues are degenerate and do not depend on $\vartheta$. The same eigenvalues are
obtained for the output states $\rho_2, \rho_3, \rho_4$. The Von Neumann entropy $S(\rho_i)$ is minimized as a
function of $\vartheta$ when the term under square root in the expression for $\lambda_{3,4}$ is maximum. The
mutual information is then maximized for equiprobable states $\pi_i$ corresponding to the minimum Von Neumann
entropy. Therefore for $\eta^2> [\eta^2(1-\mu)+\mu]^2$ the mutual information is maximal for uncorrelated states
$\vartheta=0$, while for $\eta^2< [\eta^2(1-\mu)+\mu]^2$ it is maximal for the Bell states. The threshold value
$\mu_t$ is a function of the shrinking factor and takes the form

\begin{eqnarray}
\mu_t=\frac{\eta}{1+\eta}\;. \label{threshold}
\end{eqnarray}

Therefore, for channels with $\mu<\mu_t$ the most convenient choice within the ansatz (\ref{signals}) corresponds
to uncorrelated states, while for $\mu>\mu_t$ to maximally entangled states. At the threshold value any set of
states of the form (\ref{signals}) leads to the same value for the mutual information. As an example, the behaviour
of the mutual information is plotted in Fig. \ref{Fig.1}.

\begin{figure}[ht]
\centering
\includegraphics[width=8cm]{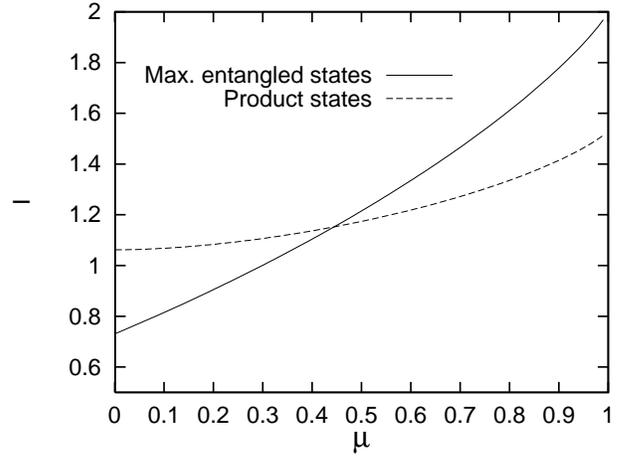}
\caption{\small Mutual information for product states and for maximally entangled states as a function of the
degree of memory of the channel, for $\eta=0.8$.} \label{cascade}
\end{figure}
%
%


It is interesting to notice that,  within the ansatz (\ref{signals}), for any value of $\mu$ the mutual information
is optimized by either maximally entangled or completely unentangled states . We have used sofar the $z$ axis as
the axis of quantisation for the system; notice that, due to the symmetry of the channel, the same results hold
also using $x$ or $y$ as the axis of quantisation.

Notice that sofar we have restricted our attention to input states of the form (\ref{signals}). We will now show
that the product states that are less deteriorated when transmitted through the channel are the eigenstates of
$\sigma_{z1}\sigma_{z2}$ or $\sigma_{y1}\sigma_{y2}$ or $\sigma_{x1}\sigma_{x2}$. This suggests that no different
choice of product signal states can achieve a higher $I_2$ than our ansatz (\ref{signals}). From Eqs. (\ref{sm})
and (\ref{cm}) it follows that the output density operator corresponding to an arbitrary input product state takes
the form

\begin{eqnarray}
\Phi(\pi)&&=\frac{1}{4}\left[\one\otimes\one + \eta(\one\otimes\sum_i \beta_{2i}
\sigma_{2i}+\sum_i \beta_{1i}\sigma_{1i}\otimes\one)\right.\nonumber\\
&&\left. +(\mu+(1-\mu)\eta^2)\sum_{i}\beta_{1i} \beta_{2i}\sigma_{1i}\otimes\sigma_{2i}
\right.\nonumber\\
&&\left. +(\mu\eta+(1-\mu)\eta^2)\sum_{i\neq j}\beta_{1i} \beta_{2j}\sigma_{1i}\otimes\sigma_{2j} \right]\;,
\label{prodotto}
\end{eqnarray}

A measure of the degree of purity of the state at the output of the channel is given by ${\mbox{Tr}}[\rho^2]$. It
is straightforward to show that for the above state we have

\begin{eqnarray}
{\mbox{Tr}}[\Phi(\pi)^2]&&=\frac{1}{4}[1+2\eta^2+(\mu+ (1-\mu)\eta^2)^2\sum_{i}\beta_{1i}^2\beta_{2i}^2
\nonumber\\
&&+ (\mu\eta+(1-\mu)\eta^2)^2 \sum_{i\neq j}\beta_{1i}^2\beta_{2j}^2]\;. \label{tr}
\end{eqnarray}

The above expression is maximised when both Bloch vectors point in the same $x$, $y$ or $z$ direction. It is
straightforward to verify that these states maximise also the fidelity, defined as ${\mbox{Tr}}[\pi \Phi(\pi)]$.
Moreover, we have numerical evidence that for any value of $\mu$ and $\eta$ the input product states that maximise
the mutual information are still of this form. Therefore, no better choice of product states leads to a higher
mutual information than the one achieved by the ansatz (\ref{signals}).

Finally we would like to point out that for input product states the mutual information $I_2(\mu = 1, \vartheta
=0)>I_2(\mu = 0,\vartheta = 0)$:

\begin{eqnarray}
I_2(\mu = 1, \vartheta =0)&=& 1 + \frac{1}{2}\{(1+\eta )\log(1+\eta
) + (1 -\eta )\log(1-\eta )\}\nonumber\\
I_2(\mu = 0, \vartheta =0)&=& \{(1+\eta )\log(1+\eta ) + (1 -\eta )\log(1-\eta )\}\nonumber
\end{eqnarray}

This is due to the fact that the correlation tensor is multiplied by a larger shrinking factor hen the noise is
collective. In other words, in the presence of perfect memory with two uses of the channel it is possible to
achieve a higher mutual information than in the case of memoryless channels even if we restrict to product states.

In conclusion, we have shown that the transmission of classical information over a quantum depolarising channel
with collective noise can be enhanced by employing maximally entangled states as carriers of information rather
than product states. We believe that this result opens new perspectives for the use of entanglement in
communications and information processing.

\section{Acknowledgements}

We would like to thank R. Jozsa for his comments. This work was supported in part by the EU under contract IST -
1999 - 11053 - EQUIP,"Entanglement in Quantum Information Processing and Communication" and by Ministero
dell'Universit\`a e della Ricerca Scientifica e Tecnologica under the project "Quantum information transmission and
processing: quantum teleportation and error correction''.



\begin{thebibliography}{99}



\bibitem{Holevo}
B.Schumacher and M.D. Westmoreland,  Phys. Rev. A {\bf 56}, 131 (1997); A.S. Holevo, IEEE Trans. Inf. Theory {\bf
44}, 269 (1998) (also quant-ph/9611023).



\bibitem{Bruss}
D. Bru\ss, L. Faoro, C. Macchiavello and G.M. Palma,  J.Mod.Opt. {\bf 47}, 325, 2000; (also quant-ph/9903033).



\bibitem{King}
C. King and M.B. Ruskai, IEEE Trans. Inf. Theory {\bf 47}, 192 (2001) (also quant-ph/9911079); C. King,
quant-ph/0103156.





\bibitem{Kraus}

K. Kraus, {\em States, Effects, and Operations: Fundamental

Notions of Quantum Theory} (Springer, Berlin, 1983).



\bibitem{Mahler}
J. Schlienz and G. Mahler {\em Phys. Rev.} A {\bf 52}, 4396 (1995).



\bibitem{Schmidt}
A. Peres, {\em Quantum Theory: Concepts and Methods} (Kluwer Academic, Dordrecht, 1993).



\bibitem{Palma}
G.M. Palma, K.-A. Suominen and A.K. Ekert, Proc. Roy. Soc. London A {\bf 452}, 567 (1996);





\end{thebibliography}
\end{document}